# Becoming a Knowledge Economy: the Case of Qatar, UAE and 17 Benchmark Countries

Osiris PARCERO[†]

James Christopher RYAN*

**Abstract**

This paper assesses the performance of Qatar and the United Arab Emirates (UAE) in terms of their achievements towards becoming knowledge-based economies. This is done through a comparison against 17 benchmark countries using a four pillars' framework comprising – (1) information and communication technology, (2) education, (3) innovation and (4) economy and regime. Results indicate that the UAE ranks slightly better than the median rank of the 19 compared countries while Qatar ranks somewhat below. Results also indicate that both countries lag considerably behind knowledge economy leaders; particularly evidenced in the innovation pillar. Policy recommendations are mainly addressed at further developing the two countries' research culture as well as improving the incentives to attract top quality researchers and highly talented workers.

**Keywords:** knowledge economy, Qatar, UAE, comparative development, information, technology, education, innovation, labor market.

---

[†] Department of Economics and Finance, United Arab Emirates University, UAE, Po Box 15551, e-mail osirisparcero@uaeu.ac.ae.

* Department of Economics and Finance, United Arab Emirates University, UAE, Po Box 15551, e-mail j.ryan@uaeu.ac.ae.

**JEL Classification:** F63, F68, O15, O25, O31

**Acknowledgements:** We would like to thank Abdul Rashid, Abdul Aziz O. Abahindy, and Shahabuddin Abdulrouf for their distinguished assistance. This paper has greatly benefited from suggestions made by Abdulnasser Hatemi Jarabad, Chiraz Labidi, David De-Meza, Henry Chappell, Melvin Damian, Mohamed Belkhir, and Wasseem Mina. Financial support from the National Research Foundation Grand # 77 UAEU-NRF is greatly acknowledged.

## 1. Introduction

Qatar and the UAE are both rich, young countries with cultural and geographic similarities. Their independence from the UK came as late as 1971. At that time Qatar and the UAE were very underdeveloped, but by 2011 they respectively rank as the 6th and 26th richest economies in the world (World Bank WDI). The two countries are highly dependent on hydrocarbons with oil and gas as percentage of export earnings reaching 91% and 69% in Qatar and the UAE respectively (Hvidt, 2013).

Moreover, the pattern of development chosen by both countries is also similar. They are highly dependent on expatriate workers – comprising 88% and 91% of the total workforce of Qatar and the UAE. Both countries also use public employment as a part of the welfare state package for their national populations (Qataris and Emiratis resp.) Not surprisingly, the large majority of nationals choose to work for the public sector where wages are sometimes "several factors higher than those elsewhere" (Forstenlechner and Rutledge, 2010, page 48). Indeed, the number of nationals working for the private sector as of 2014 was 1% and 0.5% in Qatar and the UAE (De Bel-Air, 2014; Arabian Business, 2014). However, both countries have become aware that the capacity of the public sector to absorb nationals has been largely exhausted. Consequently, localization programs, called Emiratization and Qatarization respectively, are being implemented. These programs provide incentives for the private sector to absorb higher proportions of the national labor force growth. Lastly, and as stated in the Qatar Vision 2030 and the UAE Vision 2021, both countries share the ambition to diversify their production systems away from oil and to become knowledge-based economies as well as leading regional/global players.

Knowledge and innovation have always played a crucial role in production and society in general. However, with the technological revolution and globalization process of the last few decades societies are entering a new postindustrial era, where knowledge and its rapid change has clearly become a key driver of competitiveness and social interaction. In this new scenario the role of manufacturing has been diminished to give space to a growing service sector with particular strength in the information and communication technologies. For the 10 years from 1995 to 2005 Brinkley and Lee (2007) report that knowledge-based sectors created twice as many new jobs in the United States and four times as many in Europe as low-knowledge sectors did. In the same vein, in most economically advanced countries, investments in intangibles are now as important as investments in tangibles goods. In 2005-06 the investment in intangibles was estimated to be almost half the size of tangible investment in the market sector of the Australian economy (Barnes and McClure, 2009). For the year 2005 the Finnish business intangible investments was estimated to be 9% of the GDP (Jalava et al., 2007). For the United States during the period 1995-2003 "the inclusion of intangible investment in the real output of the non-farm business sector increases the estimated growth rate of output per hour by 10–20 percent relative to the baseline case which completely ignores intangibles", (Corrado et al., 2009, page 682).

Many definitions of a knowledge-based economy are available and the World Bank (2007) suggests the following one (page 14):

"[The] knowledge economy … meaning is broader than that of high technology or the new economy, which are closely linked to the Internet, and even broader than the often-used information society. Its foundations are the creation, dissemination, and use of knowledge. A knowledge economy is one in which knowledge assets are deliberately accorded more importance than capital and labor assets, and where the quantity and sophistication of the knowledge pervading economic and societal activities reaches very high levels."

This same source considers four pillars for a knowledge-based economy, which are the main framework that guides analysis in the current paper – i.e., (1) information and communication technology (ICT); (2) education; (3) innovation and (4) economy and regime. For each of these pillars this paper assesses the performance of Qatar and the UAE through a comparison against a group of benchmark countries.

Given Qatar and the UAE's stated high ambitions, it does not make sense to benchmark them against very low performing countries. Similarly, little can be learned by comparing them against long established and highly developed economies. Thus, the approach adopted here is to benchmark them against a few relatively similar countries, which allows a better focus on country specificities as well as interesting anecdotal contrasts.

Currently, the position of Qatar and the UAE as knowledge economies in relation to other comparable countries is largely unknown. Even though a few previous assessments are available, they typically benchmark among the Gulf Cooperation Council countries (GCC) or a sub-group of Arab countries and mainly focus on education or labor market regimes – e.g., Muysken and Nour (2006); Karoly (2010); Lightfoot (2011). We differentiate from these studies by benchmarking against a larger but still comparable group of countries in relation to four pillars considered to be crucial in the development of a knowledge-based economy. In particular, we benchmark Qatar and the UAE against 17 countries – i.e., Australia, Bahrain, Chile, Costa Rica, Finland, Israel, Kuwait, Malaysia, Norway, Oman, Poland, Saudi Arabia, Singapore, South Africa, South Korea, Tunisia, and Turkey.

In addition, the very much distinctive labor market conditions in Qatar and the UAE clearly affect each of the four pillars and so deserve special consideration. As already mentioned, these conditions are: a large dependency on expatriate workers; a large majority of national choosing to work for the substantially better paid public sector's jobs; and the provision of incentives for the private sector to absorb higher proportions of the national labor force growth (through the Emiratization and Qatarization programs). The final section will be devoted to discuss the implications that these issues have on the education and mainly the innovation pillars.

The following section discusses the methodology and the choice of the benchmark countries. It is followed by four sections that respectively assess the performance of Qatar and the UAE in terms of each of the four pillars, which is done through a comparison against the performance of 17 benchmark countries. Section 7 discusses the impact of the current labor market conditions on both countries' knowledge potential. Section 8 summarizes the main results and assesses current and alternative policies. Section 9 concludes and offers policy recommendations.

## 2. Methodology

A knowledge-based economy relies on knowledge as the key engine of economic growth. Since the start of the 20th century Schumpeter (1911[1934]) identified the importance of innovation as a source of growth and as one of the main features of capitalism. Thereafter and throughout the century the idea maintained actuality, but gained momentum with the arrival of the endogenous growth theory. Endogenous growth theory holds that economic growth is primarily the result of endogenous rather than external forces, where human capital and innovation are important contributors to economic growth. Romer (1986) was one of its most representative exponents and the first in considering private incentives as a motor for the generation of knowledge (through industrial R&D) as well as introducing positive knowledge spillovers to the rest of the economy in the tradition of Arrow (1962). Many other contributions followed (Lucas (1988); Romer, 1990; Grossman and Helpman, 1991; Rebelo, 1991).

The endogenous growth approach has been shown to be successful (Barro and Sala-i-Martin, 1995; Barro, 1997; Aghion et al., 1998). However, it is difficult to single out the contribution of knowledge to total factor productivity, which is at the heart of the growth process. Indeed, Krugman (August 18, 2013) criticized this approach for the high degree of difficulty in empirically verifying it; "*too much of it involved making assumptions about how unmeasurable things affected other unmeasurable things.*" It is clear that part of the difficulties are produced by the lack of data and the fact that the concept of knowledge capital is difficult to be capture in just a few variables, many of which are highly correlated – e.g., it is not easy to completely separate it from human or physical capital.

Unlike information, knowledge involves combinations of facts that interact in intangible ways. As such it is difficult to obtain and so it can become an entry barrier to growth, though, those countries that cross the barrier benefit from the rents knowledge generates. According to Kaplinsky (2005) there are several types of knowledge rent: technological (control of cutting-edge processes or product innovativeness), human resources (possession of high human skills and know-how not easily available elsewhere), organizational (control of unique or innovative management practices), and marketing and design (with a direct connection to consumer know-how). Knowledge rents such as these do not last forever, though, and require continuous renewal.

In view of that, this paper adopts a more holistic view of the wide spectrum of factors relevant to a knowledge economy and categorize them in the four pillars mentioned in the introduction. A group of comparable countries is selected and their relative success in the four pillars is assessed through a benchmarking methodology.

The benchmarking concept in the context of organizational comparison goes back to the 1970s with the seminal study by the photocopy maker Xerox, Dattakumar and Jagadeesh (2003). Many definitions have been provided and "all imply that benchmarking is a process – i.e. a sequence of activities that involves process and assessment" Moriarty and Smallman (2009 pp.490). Kumar and Chandra (2001) suggest that benchmarking can be considered a form of "reverse engineering" where the performance goals from comparable successful organizations are assumed to be achievable and applicable to others. Alternatively, Bhutta and Huq (1999, pp.254) state that "[t]he essence of benchmarking is the process of identifying the highest standards of excellence for products, services, or processes, and then making the improvements necessary to reach those standards – commonly called best practices."

During the 1980s and 1990s corporate benchmarking techniques started to be adapted to assess the efficiency of the public services. In this context, Helden and Tillema (2005) suggest that benchmarking becomes an important substitute for the absence of market forces. They go further and suggest a theoretical approach to benchmarking by using insights from both economics as well as Oliver (1991)'s combination of neo-institutional theory and resource dependence theory.

Benchmarking has also been broadly used in the context of country comparison; see Dolowitz and Marsh (2000) for and early account of this growing phenomenon. A few exemplary implementations of the methodology stand out. First, the OECD's Going for Growth exercises, which identify five productivity-related policy priorities for each member country, OECD (2005). Second, the European Commission's Internal Market Scoreboard, which ranks member countries' performance in implementing legislation required for internal market convergence (European Commission, 2012; Ioannou et al, 2008). Third, the Knowledge Assessment Methodology (KAM) benchmarks countries' capacity to compete in what the World Bank terms the "Knowledge Economy" (World Bank, 2007).

Our paper very much borrows from the KAM initiative. This view recognizes that the conditions leading to a knowledge-based development should include an institutional regime offering the right incentives, an educated and skilled labor force, a modern and widespread information infrastructure as well as an effective innovation system. Chart 1 illustrates the relationship among these four pillars of the knowledge economy. Still, intrinsic to this view is the importance given to the fact that countries should keep a balance among the four pillars in order to create an environment that incentivizes an efficient creation, dissemination, and use of knowledge. There is also a sense of progression with, say, education being a prerequisite for innovation and hence its improvement being expected to occur earlier on in time. Moreover, this framework allows understanding of a country's strengths and weaknesses relative to other countries, hence becoming useful from the policy-making point of view. It allows pinpointing a country's problems and opportunities, revealing areas where policy attention or investments may be required to facilitate the transition to a knowledge economy.

The selection of the benchmarking countries was done through a combination of non-exclusive criteria, as follows. Being naturally resources-rich: Australia, Bahrain, Chile, Kuwait, Norway, Oman, Saudi Arabia, South Africa, UAE, and Qatar. Not being too different in terms of size (mainly population): Bahrain, Costa Rica, Finland, Israel, Kuwait, Norway, Oman, Qatar, Tunisia, the UAE, Singapore, and to a lesser extent, Australia and Chile. Close geographical or cultural distance: Bahrain, Israel, Kuwait, Oman, Qatar, Saudi Arabia, the UAE, Tunisia and Turkey. Aspiration to become knowledge-driven economy: Costa Rica, Israel, Malaysia, Qatar and the UAE, which explicitly or implicitly have expressed that intention. Already being knowledge-driven economies: Australia, Finland, Korea, Norway and Singapore. Being considered as emerging economies: Chile, Malaysia, Poland, South Africa, and Turkey (IMF, 2012). However, some countries that satisfy only some of the above mentioned criteria were not included for not fitting well with some of the remaining criteria – e.g. though being natural-resource rich, Russia and Brazil are far too large countries. Moreover, it is very reasonable not to compare Qatar and the UAE with very poor countries or those which are far from their aspirations.

The next four sections assess the performance of Qatar and the UAE in terms of each of the four pillars. This assessment is done through a comparison against the performance of the 17 benchmark countries and utilizes spider charts where a collection of indicators for each pillar can be displayed. These charts are enhanced versions of graphical illustrations proposed by the Knowledge Assessment Methodology (KAM) created by the World Bank's Knowledge for Development Program. Data availability restricts these charts to be cross-sectional, though longitudinal factors are contemplated in the discussion. The explanation of how the charts can be read is done at the time of discussing the first pillar; the meaning and source of each of the variables is indicated in the appendix Tables 1 to 4 and descriptive statistics are in the appendix Tables 5 to 8.

## 3. ICT pillar

ICT is the main infrastructure used by a knowledge-driven economy and as such it plays a similar role as railways, roads, and utilities did in the industrial era. Chart 2 shows selected ICT indicators for Qatar, the UAE and an average of the other 17 countries. Table 1 in the appendix provides the definition and source for each indicator and descriptive statistics are in the appendix Tables 5.

The first indicator in Chart 2 is *total telephones per person* – it has been shown that total telephones and also mobile phones positively affect innovation (Carayannis et al., 2013). It is measured by the axis that goes from the center to the upper vertex of the heptagon. This outer vertex corresponds to the country (among the 19 selected) having the best performance (i.e., highest *total telephones per person*). This best performing country is identified in the brackets next to the variable name, which for this variable is the UAE. On the contrary, the inner point at the center of the heptagon corresponds to the worse performing country. All the variables have been normalized to 1 in order to allow for a cardinal interpretation. Thus, the closer is the red line to the outer point the closer is Qatar to the best rather than to the worst performer. Moreover, the number in red (blue) indicates that Qatar (the UAE) ranks 3 (1), where 10 is the median rank. The other variables can be interpreted in exactly the same way.

Thus, it is clear that in terms of *total telephones per person* Qatar and the UAE show an outstanding performance. In the case of the UAE this can be explained by the fact that the predominant provider (Etisalat) is also an important actor in other countries of the region. Etisalat is a state-own company and has received longstanding support from the government in order to become a regional leader. It is currently the 12$^{th}$ largest mobile network operator in the world with operations in 15 countries across Asia, the Middle East and Africa. The local market was not excluded from this strategy and so the company achieved a large network expansion. In 2006 another state-owned company (Du) started operations and a regulator was instituted. On the other hand, Qatar broke its telecommunication monopoly when British Vodafone began operating as a second mobile telecom provider in 2009, ending a 20-year-long monopoly by Qtel. Facilitated by a period of rapidly expanding population this entrance lead to fierce competition resulting in Vodafone controlling approximately one third of mobile telephony. Not surprisingly and given that landlines are an extremely low proportion of the market, a similarly good performance is shown by both countries in terms of the number of mobiles per capita.

In sharp contrast with the previous indicator, Qatar and the UAE show a substantially lower performance in terms of *computers per person*, ranked $16^{th}$ and $10^{th}$ respectively. It can be claimed that from the perspective of a knowledge economy the number of computers is a better indicator than the total telephones – the former is more linked to productivity (Cardona, 2013). This conjecture is very much confirmed when looking at three of the knowledge economy leaders (Finland, Singapore and South Korea). They are quite on the top positions in terms of computers per capita ($1^{st}$, 3th and $7^{th}$ resp.) while in a much lower rank in terms of *total telephones per person* ($5^{th}$, $6^{th}$ and $12^{th}$ resp.)

The UAE shows a very good performance in terms of *international internet bandwidth* and *internet users per person* – ranked $4^{th}$ in both. However, it is not doing well in terms of *fixed broadband internet tariff* ($16^{th}$). This is explained by the fact that the two internet service providers do not share each other's networks, which ultimately implies they exert limited tariff competition (Implacable adversaries: Arab governments and the Internet, 2010). On the other hand, Qatar shows a very low performance in the three indicators – it ranks $14^{th}$, $18^{th}$ and $19^{th}$ in terms of *international internet bandwidth, internet users per person*, and *fixed broadband* *internet tariff* respectively. This is the case because no competition was introduced in Qatar's internet services, which is in clear contrast to what happened in its mobile market. Indeed, a royal decree has given Q-Tel a total monopoly in the internet market until 2013. Not surprisingly, this results in less provision and higher tariffs than would be expected under a competitive environment.

Chart 2 shows that in terms of *availability of e-government services* both countries are around the median rank (Qatar's 9$^{th}$ and the UAE's 10$^{th}$) and also have a better performance than the benchmark-countries' average. In addition, in terms of *e-government index* (a measurement of its quality) both countries again show a better performance than the benchmark-countries' average, with the UAE improving its rank (7$^{th}$) and Qatar moving down a few steps (13$^{th}$). The advantages of a modern E-government are well documented (Ambali, 2010; Torres et al., 2005), though, arguably this is of less importance in countries with no taxation and where many of the decisions are of the top-down style. Natural resource richness and its induced lower taxation are typically associated with lower level of government accountability (McGuirk, 2013), which negatively affect the E-government services (Navarra and Cornford, 2007). Keeping this in mind, the relative performance of the two countries seems quite satisfactory. Moreover, the UAE E-government services grew much faster than the world average in the first decade of the millennium (Farooquie, 2011) and to a slightly lesser extent this is also the case for Qatar.

Finally, by taking the arithmetic mean of the rankings of the seven indicators in Chart 2 for the UAE and Qatar respectively we arrive at the average ranks of Qatar and the UAE – 13 and 7.4 respectively. This indicates that the UAE's overall performance in terms of the ICT pillar is very good, while Qatar's is below the median rank. This is despite the fact that Qatar has a very large proportion of its population concentrated in and around its capital (Doha), which has the potential of making the provision of ICT more economical.

## 4. Education pillar

Education is the foundation of a knowledge-driven economy and as such its improvement is expected to be observed from the very first stages of economic development. Education facilitates understanding and communication, which is a crucial component of any innovation system. Furthermore, the majority of nowadays ideas and inventions are produced in educated environments (knowledge clusters) where scientific or complex communication is required (Buesa et al., 2010; Carayannis and Campbell, 2012). Chart 3 shows the selected indicators for the *education pillar* and Table 2 in the appendix provides the definition and source for each of them and descriptive statistics are in the appendix Table 6. An arithmetic mean of the rankings of these indicators for the UAE and Qatar respectively shows that the UAE's overall performance in the education pillar is slightly above the median rank (an average ranking of 9.2) while Qatar's is below (an average ranking of 13.4).[1] These findings are not very different from the two countries' performance in terms of the indicator *average years of schooling* shown in Chart 3. That is, the UAE is slightly above the median rank (ranked 11th) as well as very close to the benchmark-countries' average; while Qatar ranks 16th and is well below this average. The low performance of Qatar is not new (see Karoly, 2010; Berrebi et al., 2009) and can be partially explained by the high proportion of low skill expatriate workers among its population. This proportion is higher than in the UAE because Qatar's building of infrastructure occurred at a slightly latter stage. For the year 2010, for which the measure was taken, the UAE had practically exhausted its infrastructure expansion and was moving into a new stage.

---

[1] However, these average rankings may be slightly overstated because they are calculated under the assumption that the six countries for which there is no data for the two PISA test would have been the worst performers in these tests.

More extreme results are obtained when looking at different indicators of government expenditure on education. Chart 3 reports *public spending on education as % of GDP*, for which Qatar and the UAE rank 19th and 18th respectively. However, this indicator may underestimate the real performance of the two countries because of their high GDP per capita. An alternative measure, not reported in Chart 3, is the *public spending on education as total % of government expenditure*, which shows quite the reversed results for the UAE, which is only second to Oman. However, Qatar's low performance is also observed under this alternative measure.

In terms of *gross secondary enrollment rate* the UAE is above the median rank (9th) and well above the benchmark 17 countries' mean while Qatar's rank is only 16th. Qatar's poor performance may help to explain why it ranks last (12th) in the PISA assessments for the 15-year-olds' math and science literacy. This program is a worldwide study by the Organization for Economic Co-operation and Development (OECD) in member and non-member nations. On the other hand the UAE's PISA scores, though around the median rank (7th) in both cases,[2] are lower than the benchmark-countries' average. This relative low performance is certainly influenced by two facts, that Finland and Singapore are among the world leaders and that the five countries that did not participate in the PISA are relatively low performers in terms of *gross secondary enrollment rate*.

A more deceptive scenario is evident for Qatar and the UAE when looking at the *gross tertiary enrollment rate*, where their rankings go down to 19th and 14th respectively. There are two explanations for both countries low tertiary enrolment relative to secondary enrolment. First, a large proportion of the low skill expatriates, who typically come for a two-year period contract, are in the age of attending tertiary education. Naturally, low skill workers are more common among the youth and in the case of Qatar and the UAE this is reinforced by the fact that a substantial number of low skill expatriates are not allowed to bring their families to the country, which acts as a deterrent for older non-skill workers.[3] Second, among the children's expatriates of higher education-age who decide to stay in the country there is a significant proportion that prioritizes working over studying. Correspondingly, many of those who decide to continue in education opt for doing so in their countries' of origin. Private universities are expensive in both Qatar and the UAE and many of the expatriates' native countries offer more economical options. In the case of expatriates from the West a more common reason is the seeking of higher educational quality.[4]

---

[2] Notice that Bahrain, Kuwait, Oman, Malaysia and South Africa did not participate in the PISA.

Having said that, the UAE is still a leader in the region in terms of tertiary education and both Qatar and the UAE have announced ambitious plans to become the regional base for world-class higher education. As a result, by 2009 the UAE hosted over 40 international branch campuses, which represents almost a quarter of all international branch campuses worldwide (Becker, 2009). Qatar has invited selected international institutions to establish in one campus known as Education City and some very good universities have been attracted – i.e., Virginia Commonwealth University, Weill Cornell Medical College, Texas A&M University, Carnegie Mellon University, Georgetown University, and Northwestern University.

Finally, in terms of *tertiary school completion (% of pop 15+)* the UAE ranks 4$^{th}$ while Qatar is the median rank and very close to the mean of the benchmarked 17 countries. This clearly contrasts with their lower performance in terms of the *gross tertiary enrollment rate* shown above. This apparent contradiction can be explained by the fact that in Qatar and the UAE many expatriates obtain their degrees before coming to the country and so they do not depend much on the graduates produced by their own universities. As such, Qatar and the UAE continue to rely on imported professionals to support their innovation system, which has been already suggested as a concerning issue by Muysken and Nour (2006).

---

[3] In the UAE this is the case if the wage is below Dh3,000 and the accommodation is provided by the employer, or Dh4,000 for those who pay for their own accommodation.
[4] The dissatisfaction with the quality and cost of private education in both countries was highlighted in the early study by McGlennon (2006), though things have improved since then.

## 5. Innovation pillar

A good innovation system consists of an interconnected array of universities, research centers, firms, consultants, and other organizations that create, assimilate and adapt knowledge. The innovation pillar can be considered the most output-like pillar, particularly when it effectively manages to integrate knowledge into the production system. Chart 4 shows the selected indicators, Table 3 in the appendix provides the definition and source for each of them and descriptive statistics are in the appendix Tables 7.

The indicator *S&E journal articles per capita* shows a very low performance for the UAE and even worse for Qatar, ranked $14^{th}$ and $16^{th}$ respectively. Indeed, both countries are closer to the worse performer than to the leader (Finland). This is a result of the fact that until recently the universities focused on teaching and were behind in terms of research (Adams et al., 2009). A lot of improvement has been achieved in the last years, but there is still a long way to go.

In terms of *intellectual property protection* the UAE is doing well (ranked $5^{th}$) while Qatar is just below the median rank (ranked $9^{th}$). It has been argued that stronger patents protection leads to more innovations (Arrow, 1962), though possibly in a non-monotonic concave relationship (Lerner, 2004; Panagopoulos, 2011; Panagopoulos, 2009). Unfortunately, the relatively good intellectual property protection framework of the two countries did not translate yet into intellectual property creation, as measured by the *patents per capita granted*. This is particularly evident for the UAE (11$^{th}$), but also true for Qatar (12$^{th}$). The two countries' performance show that they are still extremely far from the leader, which for both indicators is Finland. Similar results are obtained when looking at the *private sector spending on R&D*, (UAE 9$^{th}$ and Qatar 12$^{th}$). Again, in this indicator both countries are quite far from the leader (Finland), which has encouraged private spending in R&D through a long history of collaboration-targeted public funding policy (Czarnitzki et al., 2007; Schienstock and Hämäläinen, 2001).

Another important indicator is *university-company research collaboration*, which has the potential to break the obstacles separating the intrinsic subcultures of universities and firms (Carayannis, 2014). Arguably, this becomes all the more important in countries like Qatar and the UAE with large expatriate labor force and cultural diversity. Yet, for this indicator the ranks showed by Qatar and the UAE (9$^{th}$ and 13$^{th}$ resp.) fall short of what one would expect based on this claim.

A surprisingly excellent performance is shown by Qatar and the UAE in terms of *firm-level technology absorption*, ranking 3$^{rd}$ and 1$^{st}$ respectively. On the one hand, this good performance is partially explained by the two countries' leadership encouragement of the adoption of new technologies (Oprescu, 2011). This is evident, for instance, when looking at the emphasis the policy makers place on the importance of use of technology in teaching (Al-hawari and Mouakket, 2010) – e.g., both countries enjoy a widespread use of smart blackboards in state-owned universities. Moreover, in the UAE's main federal universities each student is equipped with an I-pad, which have recently complemented and even replaced laptops. On the other hand, technology adoption is a very much capital intensive activity and as such is higher in richer and more open economies. Both Qatar and the UAE are among the richest countries in the world and their economies are relatively open in terms of both imports and *foreign direct investment* (FDI).

Chart 4 shows that in terms of *FDI inflows as % of GDP* Qatar and the UAE rank 5th and 3rd respectively. Among the reasons that make the two countries attractive to multinational firms are lack of taxation, low labor costs and flexible labor market conditions (Parcero et al., 2014). Being located in a highly conflicted region has its disadvantages, but it may have also provided some benefits. It is not uncommon to see foreign companies choosing Qatar or the UAE as their location when their intention is to find a hub through which to supply the Middle East region.

The last innovation indicator in Chart 4 is the *high technology % of the manufactured exports*, which is perhaps the clearest example of an innovation output and, as such, one of the ultimate targets for a country wishing to be considered as a knowledge-based economy. In other words, this is a good indicator that a country has taken the knowledge into the production sphere and that this knowledge is an important component of the country's exports. Unfortunately, in terms of this indicator, the UAE ranks $13^{th}$, though it is better than the $19^{th}$ ranked Qatar. Indeed, both countries are extremely far from the leader (Singapore) as well as well below the benchmark countries' average.

## 6. Economy and regime pillar

The *economy and regime pillar* acts as a facilitator for the efficient allocation of resources, stimulates entrepreneurship and the creation of knowledge as well as its dissemination. It covers a diversity of issues and policy areas ranging from aspects of the business environment, finance and banking, macroeconomic framework, regulations, governance and institutional quality. Chart 5 shows the selected indicators for this pillar, Table 4 in the appendix provides the definition and source for each of them and descriptive statistics are in the appendix Tables 8.

In terms of *regulatory quality* it is clear that both Qatar and UAE performance (ranked 11$^{th}$ and 12$^{th}$ resp.) are below the benchmark countries' average and the median rank and indeed quite far from the leader Singapore. The UAE does even worse in terms of *rule of law* (ranked 14$^{th}$); while Qatar is better than the median rank (ranked 7$^{th}$). On the one hand, smaller countries find it easier to keep the *rule of law*, which may explain the advantage that Qatar shows over the UAE. However, it does not help to explain the disadvantage of the latter relatively to the benchmark countries' average and the median rank.

In terms of the *corruption perceptions index* Qatar and the UAE are substantially better than the median rank (ranked 7$^{th}$ and 6$^{th}$ resp.) as well as the benchmark countries' average. Indeed, similar results are obtained if using the corruption index from the WEF Global Competitiveness Report 2010. This good performance is the result of the steady improvements shown by both countries, mainly from 2003 to 2006.

Even though in terms of *tariff & nontariff barriers*, both countries are slightly below the median rank (Qatar 13$^{th}$ and the UAE 12$^{th}$), they are very close to the benchmark countries' average and closer to the best rather than the worst performer. The literature shows that smaller countries are typically more open to trade (for instance Alesina and Wacziarg, 1998).

Both Qatar and the UAE take the lead for the variable *intensity of local competition*, ranking first and second respectively, and their performance is well above the benchmark countries' average. The *intensity of local competition* is a perception measure resulting from a survey done by the WEF Global Competitiveness Report 2010. Two explanations for their very good performance are in place. First, there is the fact that business is tax free in both countries. Second, this high competition is possible thanks to the large expatriate population who make partnership with national sponsors (Qataris and Emiratis resp.) to open businesses outside the free economic zones, mainly involving the provision of services. These sponsored businesses require 51% local ownership. The lack of good outside options for these expatriates makes them willing to participate in these highly competitive businesses. As a consequence, intensity of competition for these services is high and prices are low.

Even though Qatar and the UAE rankings in terms of *days to start a business* ($12^{th}$ and $13^{th}$ resp.) are slightly below the median rank, they are much closer to the leader than to the worse performer and also better than the benchmark countries' average. This good performance seems to be in line with the two countries' business-friendly traditions. However, it may be overstated if it is kept in mind that outside the free zones only nationals or sponsored expatriates can start a business. On the other hand, in terms of the indicator *time required to enforce a contract* the UAE (ranked $8^{th}$) appears very close to both the benchmark countries' average as well as half distance between the worse and best performer. However, Qatar (ranked $16^{th}$) is quite below the benchmark countries' average and so in need of further improvement of this judicial process. Enforceability of contracts is crucial for business; long periods of disputes create costs and uncertainty that ultimately discourage investment.

The *soundness of banks* is the worst indicator for the UAE in this pillar (ranked $15^{th}$) and the second worst for Qatar (ranked $14^{th}$). The fact that both countries are so wealthy and with abundant capital for potential lending does not help in explaining their low performance. However, three factors may be important determinants of their relatively low ranks. First, the banking industry is divided between conventional banking and Islamic banking. Islamic banking is a relatively new field of banking and as such still developing. It may easily be the case that there are learning economies in place and that the industry is not yet reached an optimum level. A second reason is their extremely large expatriate population who, under the impossibility of acquiring permanent residence, is more prone to leave the country at any time. This makes the banks more reluctant to make loans as it has happened that people leave the country without honoring them. This fact alone substantially reduces the market size as well as increases the moral hazard and adverse selection problem. Last but not least, the banking sector in both countries is at the forefront of the private sector labor nationalization programs (Qatarizarion and Emiratization), which basically requires the banks to comply with a minimum quota of national employees, who are more expensive and less productive (Toledo, 2013).

**7. The role of the labor market conditions**

Clearly, the functioning of the labor market is a very important element in this discussion and is also the main aspect in which there is space for both Qatar and the UAE to make improvements. Not surprisingly, the already mentioned distinction between nationals and expatriates is also relevant when analyzing the labor market. Thus, our analysis is made through these lines and starts by looking at the labor market for nationals.

Qatar and the UAE are perhaps the two most extreme welfare states when it comes to the treatment of their own nationals (Toledo, 2013). Citing a study conducted by Zayed University in Dubai, Bloomberg Business News (October 3, 2007) reported that the average Emirati male receives around $55,500 a year in total transfers. Not surprisingly then, the two countries face the lack of incentives typical in welfare states. Moreover, the extremely generous wages for the nationals in the public sector demotivate them from moving to the private sector. Both countries' governments have tried to change this through the implementation of what is called Qatarizarion

and Emiratization programs respectively. Each of these programs imposes, for selected sectors, a minimum expatriate/national personnel ratio; often called 'quota system'. It started with the banking sector in the UAE and with the oil and gas industry in Qatar, but in both countries it has expanded to other sectors. Original deadlines were set for companies to meet with these ratios and even though they have been repeatedly extended, companies get constant pressure to improve the ratio – see Kamrava (2009) and Toledo (2013) for more details.

On the one hand the quota system is beneficial because it helps nationals acquire the type of skills needed in the private sector. Given the high reservation wage of nationals this system only affects high-skill jobs. Typically, the nationals get de facto trained by expatriates and quickly end up in a higher position than their trainers or even replacing them. This benefits the local economy as nationals have a close to nil probability of leaving the country, as shown by the good performance of both Qatar and the UAE in terms of brain-drain indicator (WEF survey, 2010). Moreover, it is believed that this program helps reduce the national population's stigma associated with private sector employment - working in the public sector was and still is considered to provide higher status in a culture that is focused more on prestige than performance (Forstenlechner and Rutledge, 2010; Forstenlechner et al., 2012).

On the other hand, the quota system is costly for private firms because nationals have a high reservation wage, which is highly determined by comparable pay scales in the public sector. A recent survey by GulfTalent.com (2012) finds that around 25% of the UAE nationals expected a monthly salary of AED 25,000 (USD 6,800) in their first job, while around 10% expected AED 35,000–50,000 (USD 9,500–13,600). According to this same study starting salaries in the private sector are about one-third of that expected by UAE nationals. This is further aggravated by perceptions that nationals are considered less productive (Toledo, 2013), which in part may be influenced by laws that make extremely difficult to fire them. The significance of the cost imposed by the quota system may be more vividly seen through anecdotal reports that some companies pay nationals a mutually agreed amount for just adding them to their staff rosters, allowing the company to comply with required quotas while avoiding the payment of the high reservation wage for the full employment of a national.

Turning onto the expatriate labor market a relevant question is how do Qatar and the UAE attract talented expatriates? The simple answer is that both countries offer a relatively high quality of life for high skill workers, though this is more obvious for the UAE (The Economist Intelligence Unit, 2014). On the one hand, this is the case because of natural reasons such as hot weather and closeness to the sea. On the other hand, they have a widespread use of English, are free of taxes, have a relatively low cost of living, have a wide range of options in terms of entertainment as well as a relatively good quality of primary (and to a lesser extent high) English schools. Last but not least, it is also the case that both countries attract them by the offering of substantially better payment packages than Europe, the USA and other developed countries. This is a consequence of the scarcity of skilled workers among the national population (Doha News, June 29, 2014; Expatnetwork, 2014).

## 8. Results and Discussion of Current and Alternative Policies

The analysis in sections 3 to 5 shows that in the overall for the four pillars the UAE ranks slightly above the median rank of the 19 benchmarked countries while Qatar lies somewhat below. In terms of the economy and regulatory pillar both Qatar and the UAE performance is relatively good by being around the median ranking and the mean of the selected benchmarking countries, with Qatar slightly over-performing the UAE. However, in both countries there is space for further development in terms of *soundness of banks* and the *regulatory quality*. Moreover, Qatar still needs to reduce the *time required to enforce a contract* and the UAE needs improvements in terms of *rule of law*.

The UAE shows a very good performance in the ICT pillar achieving an average rank of 7.4 and being outstanding in indicators such as *telephones and internet users per person*. On the contrary, Qatar average ranking (13) shows that further improvements are needed, though heavy investments have been done, and more expected, in the run up to the 2022 World Cup. Its best performance is in terms of *total telephones per person* for which it ranks $3^{rd}$.

In terms of education, the UAE overall performance is slightly above the median rank (avg. ranking of 9.2) while Qatar's is below (avg. ranking of 13.4). Qatar's performance is even worse when looking at its distance to the best performers, as evident in the small area covered by the red pentagon in Chart 3. Moreover, particularly noticeable is the low performance of both the UAE and Qatar in terms of public spending on education as % of GDP (ranking $19^{th}$ and $18^{th}$ respectively). Having said that, it should be mentioned that in the last few years the UAE made significant efforts to improve its educational system and so better records can be expected in the near future. To a lesser extent this is also the case for Qatar.

In terms of the innovation pillar the UAE does well in some indicators, and to a lesser extent also Qatar. Moreover, an average of the rankings of these indicators shows that in the innovation pillar the UAE and Qatar respectively rank slightly above and below the average; i.e. 10.6 and 8.6. However, it is clear that both countries did not achieve the stages of the leaders, particularly in the most critical indicators – *journal articles per capita*, *patents per capita*, and *high technology % of the manufactured exports*. For instance, in terms of both *journal articles* and *patents per capita*, Qatar and the UAE are quite far from achieving this level at least in the following 15 or 20 years. This is evident when looking at Finland, whose current innovation activity is based on a continues increase since the mid-1990s, as measured by scientific publications, patents indicators or other innovation scoreboards (Dahlman et al., 2006).

Our assessment of the labor market conditions in Qatar and the UAE rises the issue that the dependence on generous payment packages deserves to be more closely scrutinized. In other words, in the long run, the two countries should aim at downplaying this way of attracting talents. There is evidence to suggest that intrinsic motivators are more important in the performance of knowledge workers than extrinsic motivators (Ryan, 2014). Arguably, talented people who are mainly attracted by high wages are likely to be less committed with the country than those attracted by professional development reasons or those looking for a longer term residency. Any improvement in these directions will allow the attraction of talents at a lower cost.

Two other underexploited and related options for attracting talents should be considered. On the one hand, there is the creation of a better professional development environment. Limited developmental opportunities weaken the expatriate knowledge workers willingness to remain in the country for long due to concerns over an inability to upskill (Abdulla et al., 2010). Such turnover would ultimately lead to the loss of valuable country/region-specific knowledge, which will require further time and resources to rebuild in the new incoming expatriates. Moreover, a better professional development environment complements any increase in the national skilled labor force who would also benefits from this environment. Thus, the availability of a good tertiary education system to build this skilled pool is a must. As already mentioned in the education pillar section, both Qatar and the UAE produce less graduates and technicians than the ones needed, though there have been substantial improvements in the recent years. The increase in their domestic supply would be of great help in terms of creating a flourishing environment leading to a virtuous circle where the generation of talent also results in its attraction from its regional neighbors. The backwardness of the neighboring region in this respect may play in their advantage.

On the other hand, the two countries should consider the offering of longer term residency for talented expatriates. Survey results show that longer term residency was highly valued by GCC expatriates (Naithani at al., 2010). Both Qatar and the UAE are very limited in this respect because the provision of nationality and permanent residency are extremely rare. The offering of the nationality option is highly unlikely to occur any time soon – this would be expensive if the new citizens also get entitled to the extremely generous welfare state enjoyed by nationals. A more plausible option is the offering of a longer term residency. By 2011 Qatar was considering this option in a bid to retain talent in the run up to the 2022 World Cup, but until now no progress has been made in this direction.

## 9. Conclusion

This paper compares 19 countries with special focus on the UAE and Qatar in terms of their achievement towards becoming knowledge-driven economies. From our analysis it becomes apparent that both countries have already created a good economy and regulatory environment. However, if they if Qatar and the UAE wish to realize their aspirations to be knowledge-economy leaders then significant improvements will be needed, particularly in two complementary developments. Firstly, they need to improve incentives aimed at attracting top quality researchers as well as highly talented workers. This can be achieved by the offering of permanent or longer term residency for top researchers or highly talented people; Singapore has already successfully started such a program. In addition, universities should become more research focused in order to entice top quality researchers, which requires lower teaching loads and administrative duties for faculty. For highly talented workers more free or low cost training and professional development courses should be made available or facilitated.

Secondly, it is necessary to create a genuine research culture among the nationals, where research ultimately gets the deserved status and leads to high responsibility as well as decision-making positions. This cannot be done overnight because such a culture needs to be transmitted through, at the very least, one generation. Unfortunately, in Qatar and the UAE the large majority of parents of current bachelor students do not have a higher degree. Thus, it is imperative that the seed is planted now by generating more and better quality research graduates. They will not only do the very much needed research, but also act as incubators of a culture that will nurture future researchers.

It is necessary to develop high standard research-based graduate programs within the country as well as continuing to encourage and support more nationals to study in high quality institutions abroad. This should be complemented with the provision of the right incentives for these graduates to enter research employment rather than government managerial positions. They may take the form of requiring a minimum period of research service before moving to other positions, improving the researchers' salary relative to the civil servants', providing more research grants or creating more research positions at the different government agencies.

If the right incentives are put in place, both Qatar and the UAE will be able to successfully achieve the knowledge-driven economy ultimate and most difficult stage where knowledge is constantly generated, incorporated into the production system as well as exported. On the contrary, if the two countries fail to implement the necessary changes a different trajectory would emerge. They may end up becoming fairly educated and technology adopting countries, but failing to emerge as product and service innovators as well as creators and exporters of knowledge. History have some sad examples of this type; the Soviet Union and the Democratic Republic of Germany did not manage to translate their high educational achievements into the development of new products or high technological exports.

35. Grossman, G. M., & Helpman, E. (1991). Trade, knowledge spillovers, and growth. *European Economic Review*, 35(2), 517-526.

36. GulfTalent.com (2012). Employment and Salary Trends in the Gulf, http://www.gulftalent.com/home/Employmentand-Salary-Trends-in-the-Gulf-2012-Report-33.html. (Retrieved on May 1, 2015).

37. Hvidt, M. (2013). Economic diversification in GCC countries: Past record and future trends.

38. IMF (2012). World economic outlook update: new setbacks, further policy action needed http://www.imf.org/external/pubs/ft/weo/2012/update/02/index.htm. (Retrieved on May 1, 2015).

39. Implacable adversaries: Arab governments and the Internet (2010). http://old.openarab.net/en/node/346. (Retrieved on December 25, 2011).

40. Ioannou, D., Ferdinandusse, M., Coussens, W., & Lo Duca, M. (2008). Benchmarking the Lisbon strategy. ECB Occasional Paper, (85).

41. Jalava, J., Aulin-Ahmavaara, P., & Alanen, A. (2007). Intangible capital in the Finnish Business sector, 1975-2005 (No. 1103). ETLA Discussion Papers, The Research Institute of the Finnish Economy (ETLA).

42. Kamrava, M. (2009). Royal factionalism and political liberalization in Qatar. *The Middle East Journal*, 63(3), 401-420.

43. Karoly, L. A. (2010). The role of education in preparing graduates for the labor market in the GCC countries.

44. Krugman, Paul (August 18, 2013). "The New Growth Fizzle". New York Times.
32

# Appendix

## Table 1: Indicators' definitions and sources for the *ICT pillar*

| **Total telephones per person, 2009** | Telephone mainlines + Mobile phones per person. (Development Data Platform, World Bank) |
|---|---|
| **Computers per person, 2008** | Personal computers per capita, that are self-contained computers designed to be used by a single individual. (Development Data Platform, World Bank) |
| **Availability of e-government services, 2008** | This indicator is based on a large sample group in a particular country responding to the question of whether the "online government services, such as personal tax, car registration, passport, business permit, and e-procurement are (1 = not available, 7 = extensively available) (1= low, 7 = high). (WEF Global Information Technology Report, 2008/2009) |
| **E-government index, 2012** | It is a composite measure of three important dimensions of e-government, namely: provision of online services, telecommunication connectivity and human capacity. (E-government index) |
| **Fixed broadband internet tariff, 2009** | Fixed broadband Internet access tariff is the lowest sampled cost in dollar per 100 kilobits a second per month and are calculated from low- and high-speed monthly service charges. Monthly charges do not include installation fees or modem rentals. (Development Data Platform, World Bank) |
| **International internet bandwidth, 2009** | This is the contracted capacity of international connections between countries for transmitting Internet traffic. International Telecommunication Union, World Telecommunication Development Report and database, and World Bank estimates. (Development Data Platform, World Bank) |
| **Internet users per person, 2009** | The indicator relies on nationally reported data. In some cases, it is based on national surveys (they differ across countries in the age and frequency of use they cover), in others it is derived from reported Internet Service Provider subscriber counts. (Development Data Platform, World Bank) |

**Table 2: Indicators' definitions and sources for the *education pillar***

| | |
|---|---|
| **Average years of schooling, 2010** | (15 years old and above) This variable is used as an aggregate measure of the educational stock in a country. (Barro and Lee, 2010) |
| **Tertiary School completion, total (% of pop 15+), 2010** | Percentage of population, 15+, total, completed tertiary is the percentage of people over age 15 who have completed tertiary education. (Barro and Lee, 2010) |
| **Public spending on education as % of GDP, 2009** | This consists of public spending on public education plus subsidies to private education at the primary, secondary, and tertiary levels. (Development Data Platform, World Bank) |
| **15-year-olds' science literacy, 2009** | Scores of 15-year-old students in science literacy. (OECD Program for International Student Assessment, PISA) |
| **15-year-olds' math literacy, 2009** | Scores of 15-year-old students in mathematics literacy. (OECD Program for International Student Assessment, PISA) |
| **Gross tertiary enrollment rate, 2009** | The ratio of total enrollment, regardless of age, to the population of the age group that officially corresponds to the level of education shown. (UNESCO) |
| **Gross secondary enrollment rate, 2009** | The ratio of total enrollment, regardless of age, to the population of the age group that officially corresponds to the level of education shown. (UNESCO) |

**Table 3: Indicators' definitions and sources for the *innovation pillar***

| | |
|---|---|
| **Intellectual property protection, 2010** | This is based on the statistical score on a 1-7 scale of a large sample group in a particular country responding to the question of whether intellectual property protection is strong in their country (1= weak or nonexistent, 7 = is equal to the world's most stringent). (WEF Global Competitiveness Report, 2010-2011) |
| **Patents per capita** | This is the number of patents applications granted per capital in average during the period |

| **granted by the USPTO, avg. for 2005-09** | 2005-09. (USPTO). |
|---|---|
| **University-company research collaboration, 2010** | This is based on the statistical score on a 1-7 scale of a large sample group in a particular country responding to the question of whether companies' collaboration with local universities in research and development activities in their country is (1= minimal or nonexistent, 7= intensive and ongoing). (WEF Global Competitiveness Report, 2010-2011 ) |
| **Private sector spending on R&D, 2010** | This is based on the statistical score on a 1-7 scale of a large sample group in a particular country responding to the question of whether companies spend heavily on research in their country. (1= do not spend, 7 = spend heavily relative to international peers). (WEF Global Competitiveness Report, 2010-2011) |
| **Firm-level technology absorption, 2010** | This is based on the statistical score on a 1-7 scale of a large sample group in a particular country responding to the question of whether the companies in your country are (1= not able to absorb new technology, 7 = aggressive in absorbing new technology). (WEF Global Competitiveness Report 2010-20011) |
| **FDI inflows as % of GDP, 2004-08** | Inflows of foreign direct investment in the reporting economy comprise capital provided (either directly or through other related enterprises) by a foreign direct investor to an enterprise resident in the economy. (UNCTAD and World Bank) |
| **High technology % of manufactured exports, 2011** | High-technology exports are products with high R&D intensity, such as in aerospace, computers, pharmaceuticals, scientific instruments, and electrical machinery. (Development Data Platform, World Bank) |
| **S&E journal articles per capita, 2007** | This is the variable above, weighted by million population. Sources: Thomson Reuters, SCI and SSCI; The Patent Board; and National Science Foundation, Division of Science Resources Statistics, special tabulations. (Development Data Platform, World Bank) |

**Table 4: Indicators' definitions and sources for the *economy and regime pillar***

| **Regulatory quality, 2009** | This indicator measures the incidence of market-unfriendly policies such as price controls or |
|---|---|
| | inadequate bank supervision, as well as perceptions of the burdens imposed by excessive regulation in areas such as foreign trade and business development. (Governance Indicators, World Bank) |
| **Tariff & nontariff barriers, 2011** | This is a score assigned to each country based on the analysis of its tariff and non-tariff barriers to trade, such as import bans and quotas as well as strict labeling and licensing requirements. The score is based on the Heritage Foundation's Trade Freedom score. (Heritage Foundation) |
| **Intensity of local competition, 2010** | This is based on the statistical score on a 1-7 scale of a large sample group in a particular country responding to the question of whether competition in the local markets is intense in their country. (1= limited in most industries and price-cutting is rare, 7 = intense and market leadership changes over time). (WEF Global Competitiveness Report, 2010-2011) |
| **Days to start a business, 2011** | This variable measures the duration of all procedures required to register a firm. (Doing Business, World Bank) |
| **Soundness of banks, 2010** | This is based on the statistical score on a 1-7 scale of a large sample group in a particular country responding to the question of whether "banks are generally sound" in their country. (1= insolvent and may require government bailout, 7= generally healthy with sound balance sheets). (WEF Global Competitiveness Report, 2010-2011) |
| **Time required to enforce a contract, 2010** | Is the number of calendar days from the filing of the lawsuit in court until the final determination and, in appropriate cases, payment. Source: World Development Indicators. (Doing Business, World Bank ) |
| **Corruption perceptions index, 2013** | This is a perception of index obtained through a survey asking respondents how corrupt the public sectors are seen in countries. (Corruption Perceptions Index, Transparency International, 2013) |
| **Rule of law, 2009** | This indicator includes several indicators which measure the extent to which agents have confidence in and abide by the rules of society. These include perceptions of the incidence of both violent and non-violent crime, the effectiveness and predictability of the judiciary, and the enforceability of contracts. (Governance Indicators, World Bank) |

## Table 5: Descriptive statistics for the *ICT pillar*

|  | Total telephones per 1000 people 2009 | Computers per 1000 people 2008 | Availability of e-government services (1-7) 2008 | E-government index 2008 | Fixed broadband internet access tariff (US$ per month) 2009 | International internet bandwidth (bits per person) 2009 | Internet users per 1000 people 2009 |
|---|---|---|---|---|---|---|---|
| Australia | 1,520 | 680 | 5.6 | 0.81 | 26 | 5,457 | 720 |
| Bahrain | 2,290 | 750 | 4.8 | 0.57 | 26 | 2,521 | 820 |
| Chile | 1,180 | 310 | 5.7 | 0.58 | 48 | 4,076 | 340 |
| Costa Rica | 760 | 340 | 3.5 | 0.51 | **6** | 4,333 | 340 |
| Finland | 1,710 | **790** | 5.5 | 0.75 | 39 | 17,221 | 840 |
| Israel | 1,650 | 280 | 5.4 | 0.74 | 7 | 2,003 | 500 |
| Korea, Rep. | 1,380 | 580 | 5.5 | 0.83 | 25 | 6,065 | 810 |
| Kuwait | 1,270 | 340 | 3.0 | 0.52 | 19 | 871 | 390 |
| Malaysia | 1,270 | 230 | 5.3 | 0.61 | 19 | 5,097 | 580 |
| Norway | 1,500 | 630 | 5.8 | **0.89** | 51 | **26,904** | **920** |
| Oman | 1,510 | 170 | 4.6 | 0.47 | 31 | 1,365 | 430 |
| Poland | 1,420 | 170 | 2.1 | 0.61 | 14 | 2,748 | 590 |
| Qatar | 1,950 | 160 | 5.3 | 0.53 | 55 | 2,044 | 280 |
| Saudi Arabia | 1,930 | 690 | 4.3 | 0.49 | 27 | 1,731 | 390 |
| Singapore | 1,700 | 740 | **6.4** | 0.70 | 17 | 22,783 | 730 |
| South Africa | 1,030 | 80 | 4.0 | 0.51 | 27 | 70 | 90 |
| Tunisia | 1,050 | 100 | 4.5 | 0.35 | 12 | 2,699 | 340 |
| Turkey | 1,060 | 60 | 4.1 | 0.48 | 18 | 4,323 | 350 |
| UAE | **2,660** | 330 | 5.2 | 0.63 | 41 | 13,233 | 820 |

**NOTE:** The score for the best performer in each indicator is in **bold**.

**Table 6: Descriptive statistics for the *education pillar***

|  | Average years of schooling 2009 | Tertiary school completion, total (% of pop 15+) 2010 | Public spending on education as % of GDP 2009* | 15-year-olds' science literacy (PISA) 2009** | 15-year-olds' math literacy (PISA) 2009** | Gross tertiary enroll-ment rate 2009 | Gross secondary enroll-ment rate 2009 |
|---|---|---|---|---|---|---|---|
| Australia | 12.12 | **20.6** | 4.0 | 527 | 514 | 82 | **133** |
| Bahrain | 9.59 | 4.2 | 3.0 | n/a | n/a | 51 | 96 |
| Chile | 10.18 | 10.4 | 4.0 | 447 | 421 | 55 | 90 |
| Costa Rica | 8.68 | 10.3 | 6.0 | n/a | n/a | 25 | 96 |
| Finland | 9.97 | 10.5 | 6.0 | **554** | 541 | 91 | 109 |
| Israel | 11.33 | 18.0 | 6.0 | 455 | 447 | 63 | 89 |
| Korea, Rep. | 11.85 | 16.2 | 4.0 | 538 | 546 | **100** | 97 |
| Kuwait | 6.29 | 3.4 | 3.8 | n/a | n/a | 18 | 90 |
| Malaysia | 10.14 | 4.7 | 4.0 | n/a | n/a | 36 | 69 |
| Norway | **12.3** | 11.7 | **7.0** | 500 | 498 | 74 | 110 |
| Oman | n/a | n/a | 3.9 | n/a | n/a | 26 | 91 |
| Poland | 9.87 | 9.1 | 5.0 | 508 | 495 | 71 | 99 |
| Qatar | 7.45 | 9.1 | 2.5 | 379 | 368 | 10 | 85 |
| Saudi Arabia | 8.48 | 6.2 | 6.0 | n/a | n/a | 33 | 97 |
| Singapore | 9.14 | 10.7 | 3.0 | 542 | **562** | 56 | 63 |
| South Africa | 8.56 | 0.5 | 5.0 | n/a | n/a | 15 | 94 |
| Tunisia | 7.32 | 6.7 | **7.0** | 401 | 371 | 34 | 90 |
| Turkey | 7.02 | 5.3 | 2.9 | 454 | 445 | 38 | 82 |
| UAE | 9.5 | 11.8 | 1.0 | 466 | 453 | 30 | 95 |

**NOTE:** The score for the best performer in each indicator is in **bold**. (*) The data for Oman, Kuwait and Turkey are for 2006 while for Qatar 2008 is reported. (**) Information for the following countries is not available: Bahrain, Costa Rica, Kuwait, Malaysia, Saudi Arabia and South Africa.

**Table 7: Descriptive statistics for the *innovation pillar***

| | Intellectual property protection (1-7) 2010 | Patents granted by USPTO / mil. people, avg 2005-2009 | University-company research collaboration (1-7) 2010 | Private sector spending on R&D (1-7) 2010 | Firm-level technology absorption (1-7) 2010 | FDI inflows as % of GDP 2004-2008 | High technology % of manufactured exports 2011 | S&E journal articles / mil. people 2007 |
|---|---|---|---|---|---|---|---|---|
| Australia | 5.6 | 68.9 | 5.1 | 4.1 | 5.9 | 2.8 | 11.9 | 846 |
| Bahrain | 5.2 | 0.0 | 3.3 | 2.7 | 5.2 | 11.3 | 0.2 | 64 |
| Chile | 3.7 | 1.2 | 4.2 | 3.2 | 5.3 | 7.1 | 4.6 | 105 |
| Costa Rica | 3.6 | 3.9 | 4.5 | 3.8 | 5.1 | 6.3 | 40.8 | 22 |
| Finland | **6.2** | **729.2** | **5.6** | **5.4** | 6.0 | 2.3 | 9.3 | **936** |
| Israel | 4.2 | 176.6 | 5.1 | 4.7 | 6.1 | 5.6 | 14.0 | 924 |
| Korea, Rep. | 4.1 | 151.2 | 4.7 | 4.7 | 6.1 | 0.7 | 25.7 | 381 |
| Kuwait | 4.1 | 3.6 | 3.2 | 2.7 | 5.3 | 0.1 | 0.5 | 91 |
| Malaysia | 4.7 | 5.6 | 4.7 | 4.5 | 5.5 | 3.8 | 43.4 | 30 |
| Norway | 5.6 | 277.8 | 4.9 | 4.4 | **6.2** | 1.2 | 18.5 | 867 |
| Oman | 5.3 | 0.5 | 3.9 | 3.2 | 5.1 | n/a | 2.6 | 47 |
| Poland | 3.7 | 1.1 | 3.6 | 3.0 | 4.6 | 4.6 | 5.9 | 187 |
| Qatar | 4.8 | 1.3 | 4.5 | 3.5 | 6.1 | 6.9 | 0.0 | 42 |
| Saudi Arabia | 4.8 | 0.9 | 4.3 | 4.1 | 5.6 | 5.0 | 0.7 | 24 |
| Singapore | 6.1 | 97.0 | 5.4 | 5.0 | 6.0 | **15.8** | **45.2** | 827 |
| South Africa | 4.9 | 2.5 | 4.6 | 3.5 | 5.4 | 1.6 | 5.1 | 58 |
| Tunisia | 4.4 | 0.1 | 4.1 | 3.6 | 5.4 | 5.3 | 5.6 | 74 |
| Turkey | 2.6 | 0.4 | 3.4 | 3.0 | 5.1 | 2.5 | 1.8 | 118 |
| UAE | 5.3 | 1.6 | 4.1 | 3.9 | **6.2** | 8.1 | 3.2 | 49 |

**NOTE:** The score for the best performer in each indicator is in **bold**.

**Table 8: Descriptive statistics for the *economy and regime pillar***

|  | Regulatory quality 2009 | Tariff & nontariff barriers 2011 | Intensity of local competition (1-7) 2010 | Days to start a business 2011 | Soundness of banks (1-7) 2010 | Time required to enforce a contract (days) 2010 | Corruption perceptions Index 2013 | Rule of law 2009 |
|---|---|---|---|---|---|---|---|---|
| Australia | 1.74 | 84 | 5.7 | **2** | **6.5** | 395 | 81 | 1.73 |
| Bahrain | 0.78 | 83 | 5.2 | 9 | 5.9 | 635 | 48 | 0.51 |
| Chile | 1.50 | 88 | 5.5 | 7 | **6.5** | 480 | 71 | 1.25 |
| Costa Rica | 0.53 | 85 | 5.0 | 60 | 5.9 | 852 | 53 | 0.56 |
| Finland | 1.73 | 88 | 5.1 | 14 | 6.3 | 375 | **89** | **1.94** |
| Israel | 1.09 | 88 | 5.6 | 34 | 6.3 | 890 | 61 | 0.83 |
| Korea, Rep. | 0.85 | 71 | 5.7 | 7 | 4.7 | 230 | 55 | 1.00 |
| Kuwait | 0.20 | 82 | 5.0 | 32 | 5.3 | 566 | 43 | 0.59 |
| Malaysia | 0.33 | 79 | 5.3 | 6 | 5.7 | 585 | 50 | 0.55 |
| Norway | 1.39 | 89 | 5.5 | 7 | 6.1 | 280 | 86 | 1.88 |
| Oman | 0.66 | 84 | 5.1 | 8 | 5.7 | 598 | 47 | 0.68 |
| Poland | 0.93 | 88 | 5.4 | 32 | 5.2 | 830 | 60 | 0.68 |
| Qatar | 0.62 | 82 | **6.1** | 12 | 5.5 | 670 | 68 | 0.96 |
| Saudi Arabia | 0.22 | 82 | 5.6 | 5 | 5.9 | 635 | 46 | 0.12 |
| Singapore | **1.84** | **90** | 5.5 | 3 | 6.3 | **150** | 86 | 1.61 |
| South Africa | 0.42 | 77 | 5.0 | 19 | **6.5** | 600 | 42 | 0.06 |
| Tunisia | 0.10 | 54 | 5.4 | 11 | 5.3 | 565 | 41 | 0.22 |
| Turkey | 0.31 | 85 | 5.7 | 6 | 5.6 | 420 | 50 | 0.12 |
| UAE | 0.56 | 83 | 5.7 | 13 | 5.4 | 537 | 69 | 0.52 |

**NOTE:** The score for the best performer in each indicator is in **bold**.

**Chart 1: The four pillars of a knowledge economy**

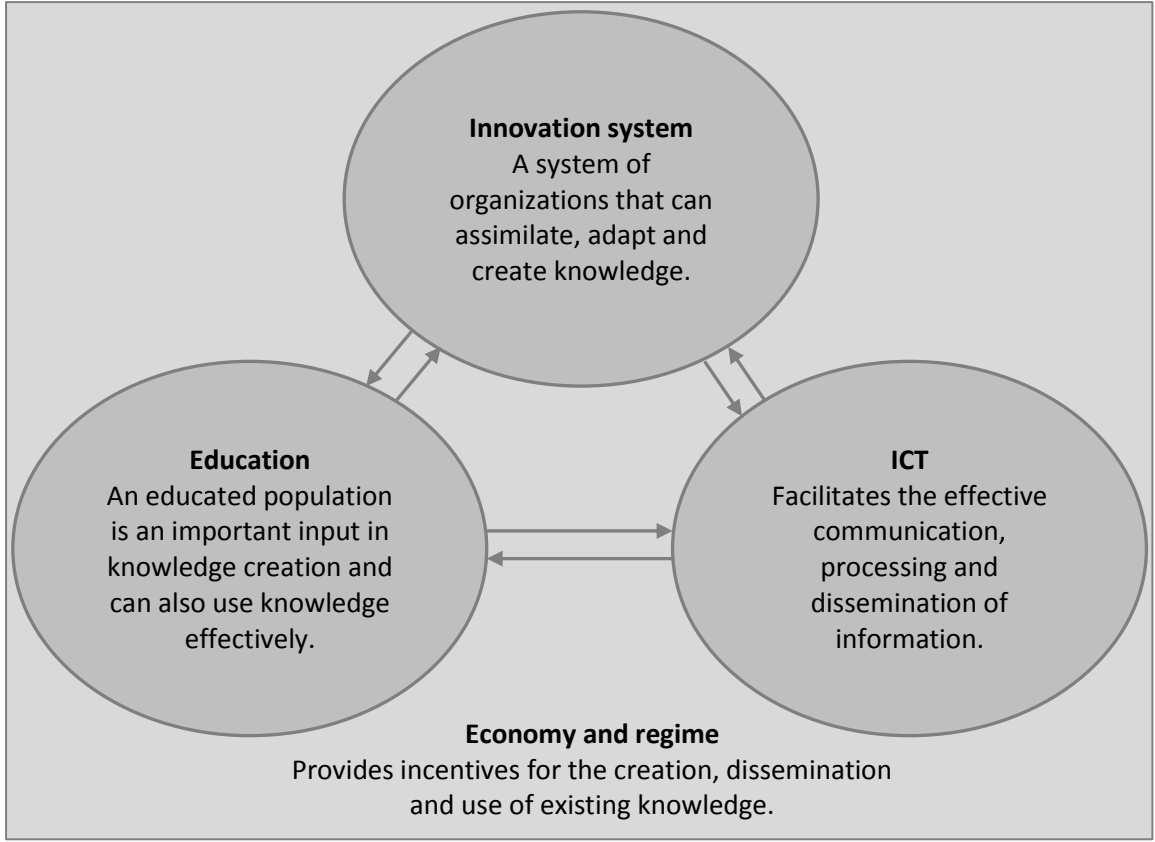

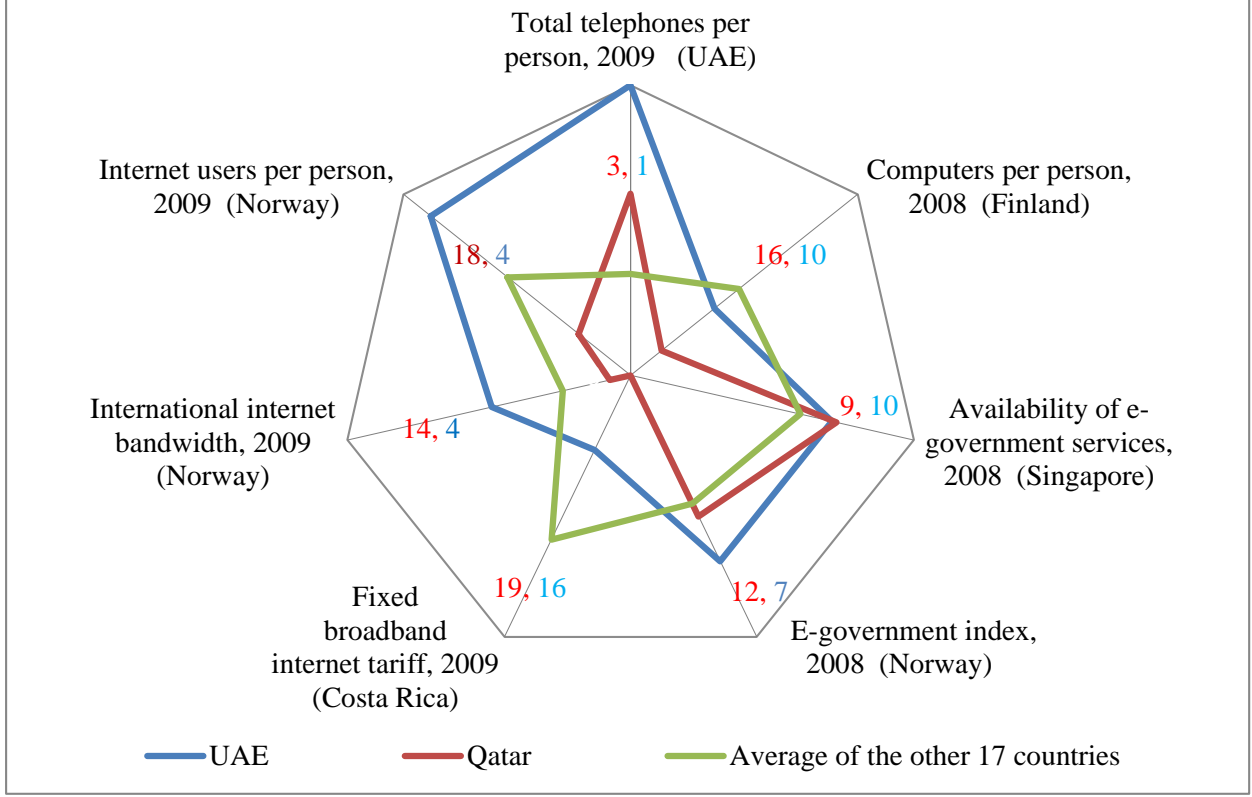

**NOTE:** The meaning and source for each of the variables are in the appendix Table 1. The explanation of how this chart is interpreted is done in the second paragraph of Section 3.

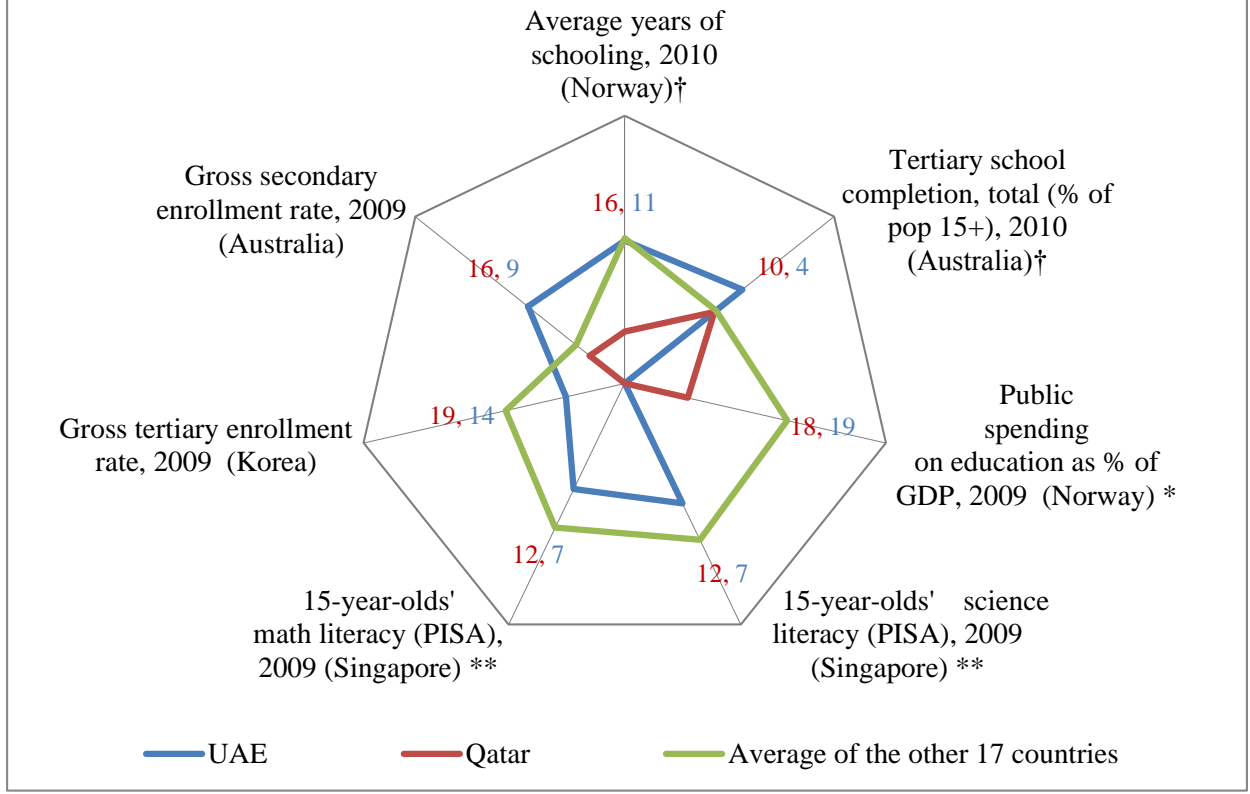

**Chart 3: Qatar and the UAE performance in terms of the Education pillar (benchmarked against the average of the other 17 countries)**

**NOTE:** The meaning and source for each of the variables are in the appendix Table 2. The explanation of how this chart is interpreted is done in the second paragraph of Section 3. (*) The data for Oman, Kuwait and Turkey are for 2006 while for Qatar 2008 is reported. (**) Information for the following countries is not available: Bahrain, Costa Rica, Kuwait, Malaysia, Saudi Arabia and South Africa. (†) Indicator not available for Oman.

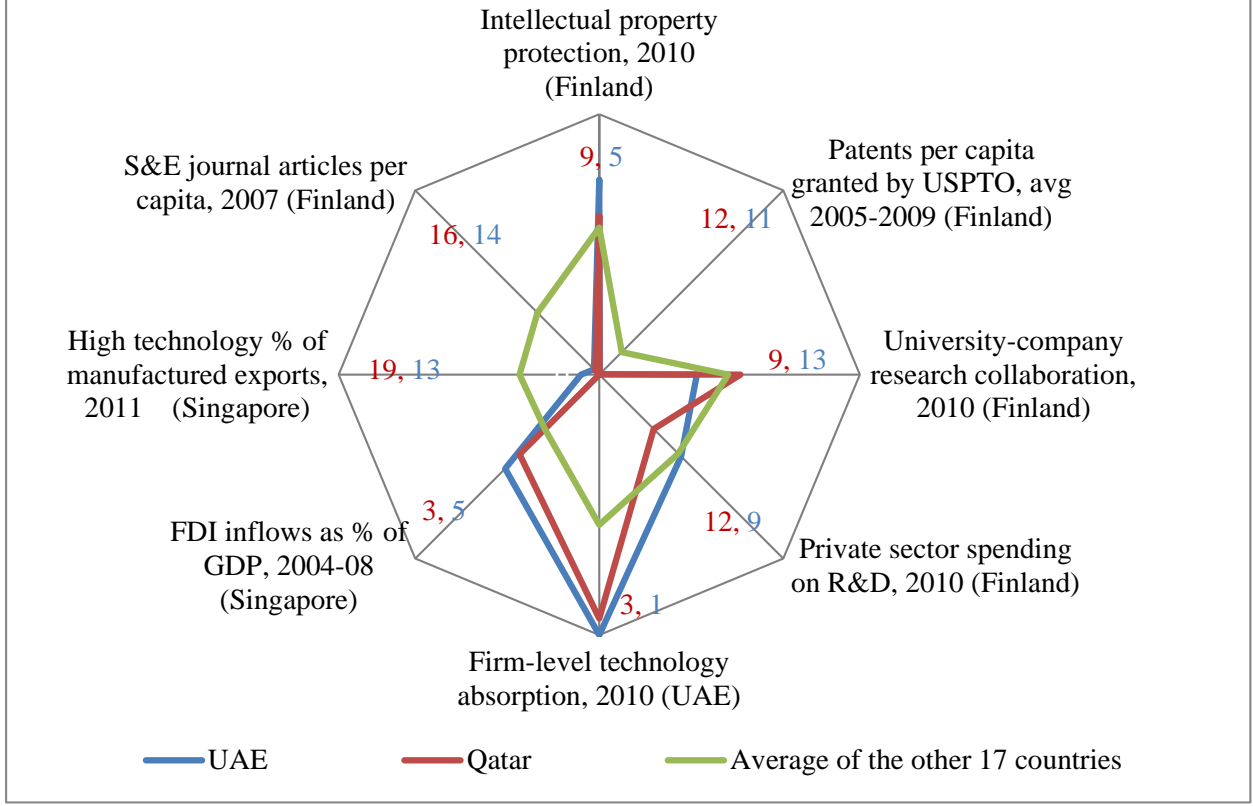

**NOTE:** The meaning and source for each of the variables are in the appendix Table 3. The explanation of how this chart is interpreted is done in the second paragraph of Section 3.

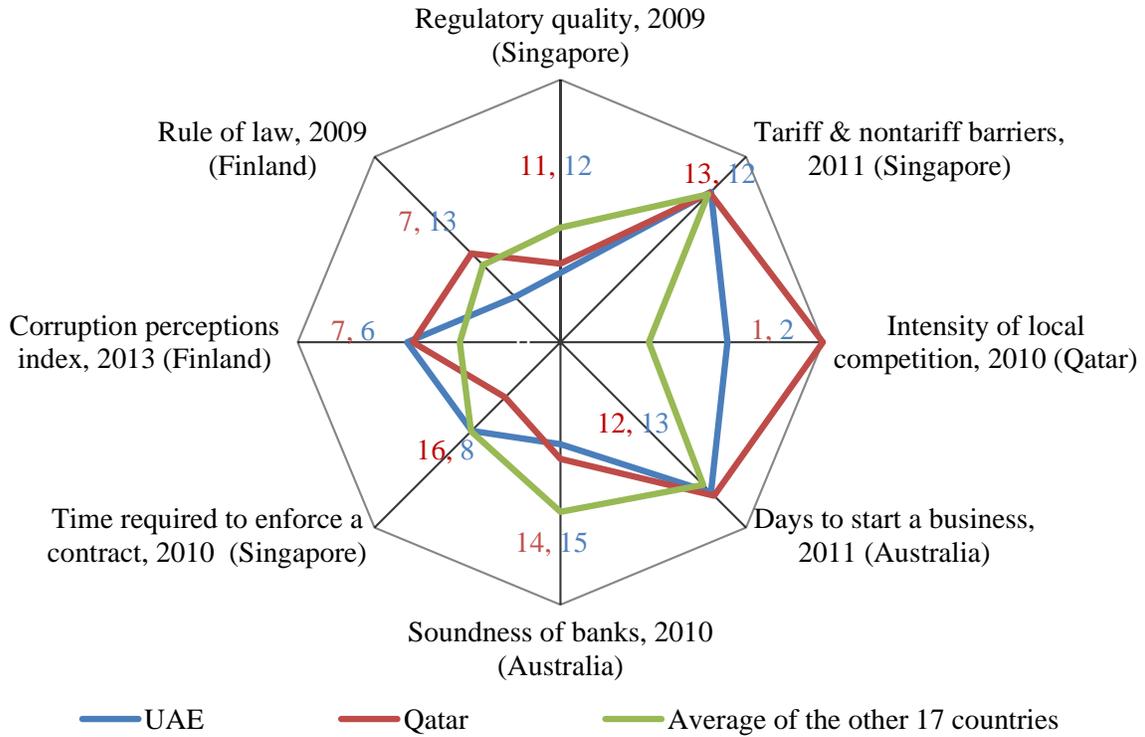

**Chart 5: Qatar and the UAE performance in terms of the *economy and regime pillar* (benchmarked against the average of the other 17 countries)**

**NOTE:** The meaning and source for each of the variables are in the appendix Table 4. The explanation of how this chart is interpreted is done in the second paragraph of Section 3.